\documentclass{svproc}
\usepackage[backend=biber, style=ieee]{biblatex}
\usepackage{url}

\usepackage[hidelinks]{hyperref}
\hypersetup{colorlinks,allcolors=blue} 

\usepackage[table,xcdraw]{xcolor}
\usepackage{makecell}
\usepackage[flushleft]{threeparttable} 

\usepackage{graphicx}

\begin{document}
\mainmatter              
\title{Impact of Data Snooping on Deep Learning Models for Locating Vulnerabilities in Lifted Code}
\titlerunning{Impact of Data Snooping}  
%
\author{Gary A. McCully\inst{1} \and John D. Hastings\inst{1} \and Shengjie Xu\inst{2}}
\authorrunning{McCully, Hastings, and Xu} 
%

\institute{The Beacom College of Computer and Cyber Sciences \\ Dakota State University, Madison SD 57042, USA\\
\email{gary.mccully@ieee.org} \\
\email{john.hastings@dsu.edu}
\and
Department of Cyber, Intelligence, and Information Operations \\ College of Applied Science \& Technology \\
University of Arizona, Tucson AZ 85721, USA\\
\email{sjxu@arizona.edu}}

\maketitle              

\begin{abstract}
This study examines the impact of data snooping on neural networks used to detect vulnerabilities in lifted code, and builds on previous research that used word2vec and unidirectional and bidirectional transformer-based embeddings. The research specifically focuses on how model performance is affected when embedding models are trained with datasets, which include samples used for neural network training and validation. The results show that introducing data snooping did not significantly alter model performance, suggesting that data snooping had a minimal impact or that samples randomly dropped as part of the methodology contained hidden features critical to achieving optimal performance. In addition, the findings reinforce the conclusions of previous research, which found that models trained with GPT-2 embeddings consistently outperformed neural networks trained with other embeddings. The fact that this holds even when data snooping is introduced into the embedding model indicates GPT-2's robustness in representing complex code features, even under less-than-ideal conditions.
\keywords{Machine Learning, Data Snooping, Buffer Overflows, RoBERTa, GPT-2, LLVM, word2vec}
\end{abstract}
\definecolor{lgreen}{rgb}{0.78, 0.99, 0.78}
\definecolor{lred}{rgb}{0.99, 0.78, 0.78}
\section{Introduction}
The use of closed-source software is common among most organizations. Applications from companies that produce closed-source software, such as Microsoft Windows, Adobe, and Oracle~\cite{5428450}, are widely used in various industries worldwide. Thus, vulnerabilities within these popular closed-source products can affect systems on a large scale. For example, WannaCry ransomware exploited a vulnerability in the Windows Server Message Block (SMB) protocol, infecting over 200,000 systems across several countries and causing an estimated \$8 billion in losses~\cite{9329407}. By definition, closed-source software does not grant organizations access to its source code, leaving companies reliant on the software manufacturer to ensure that the product is free from exploitable vulnerabilities that could impact the organization negatively. Sadly, based on worms such as WannaCry~\cite{9329407}, Sasser, Code Red, and Slammer~\cite{4483668}, we know that these software companies have repeatedly failed to create software devoid of these types of vulnerabilities.

Identifying vulnerabilities in closed-source software is challenging, as organizations are provided with only compiled binaries to deploy on their systems, making it difficult to detect vulnerabilities without access to the source code. To address this issue, many researchers are turning to machine learning techniques to pinpoint vulnerabilities within compiled code~\cite{RN9,schaad2022deeplearningbased,RN67,9948365,gallagherllvm}. \cite{mccully1} and \cite{mccully2024}, in particular, demonstrate promising results.

Given the criticality of detecting vulnerabilities in compiled code, accuracy in detection is important. However, underexplored design flaws such as data snooping can have unintended negative consequences on model performance. Unfortunately, data snooping is a widely overlooked risk. Indeed, \cite{arp2022and} found a prevalence of data snooping within current cybersecurity research, based on their analysis of 30 papers published over ten years in the top four security-focused conferences. 
Surprisingly, they found that seventeen of the papers contained data snooping design flaws, five papers had partial data snooping issues, 
four papers had an indeterminate classification, and only four of the 30 papers definitively did not contain this flaw.

With the widespread prevalence of this issue, it is crucial to understand its implications for cybersecurity applications, particularly in machine learning-based vulnerability detection. This paper examines the impact of data snooping on deep learning models used for vulnerability detection in lifted compiled code in the context of \cite{mccully1,mccully2024}. These studies served as a baseline for this project, as they involved several embedding models, including more advanced unidirectional and bidirectional transformer-based models (e.g., GPT-2, RoBERTa). Furthermore, the details in these projects were sufficient to allow for easy modifications to introduce the necessary conditions for this research. By introducing controlled instances of test snooping into the training process of embedding models, including GPT-2, RoBERTa, and word2vec, this study aims to assess how incorporating data used to train and validate neural networks into embedding training influences model performance. The goal is to determine the extent to which data snooping influences the generalization ability of neural networks in vulnerability detection.

\section{Data Snooping}\label{sec:datasnooping}
Data snooping is a design flaw that occurs when machine learning models are trained with information that is inaccessible in real world situations~\cite{arp2022and}. 
Although data snooping can appear in three forms: test snooping, temporal snooping, and selective snooping~\cite{arp2022and}, test snooping is the focus of the current study,  with brief analyses of the other forms provided in Section \ref{sec:tempselsnooping}. \textit{Test snooping} occurs when the dataset intended for the final validation of the model is used for other model development tasks~\cite{arp2022and}, which compromises the role of the test dataset as an impartial tool for evaluation. Test snooping can be further divided into four subcategories that are listed in Table \ref{tab:test_snooping}.

\begin{table}[h!]
\centering
\caption{Test Snooping Subcategories (\textit{Source:} \cite{arp2022and})}
\begin{tabular}{|p{0.2\columnwidth}|p{0.70\columnwidth}|}
\hline
\textbf{Category} & \textbf{Description} \\ \hline
Preparatory Work & Features are selected using all data samples, including those intended for model validation, thereby introducing knowledge that would not be available in a true predictive scenario. \\ \hline
K-fold Cross-validation & Test datasets are indirectly used to tune hyperparameters through techniques such as k-fold cross-validation, which compromises the test set’s role as an unbiased measure of model performance. \\ \hline
Normalization & Techniques such as tf-idf are performed on the entire dataset before the dataset is split into training and validation sets. \\ \hline
Embeddings & Embeddings are calculated using all samples, including the validation dataset. \\ \hline
\end{tabular}
\label{tab:test_snooping}
\end{table}

A review of \cite{mccully1} and \cite{mccully2024} reveals an absence of test snooping within the methodology. The following reasons support this conclusion:
  \begin{itemize}
    \item Feature selection was based on preprocessing steps that focused solely on cleaning up the lifted LLVM code without bias toward features specific to any group of samples. 
    \item Hyperparameter selection was performed without the use of k-fold cross-validation.
    \item Additional normalization techniques were not applied to the datasets beyond preprocessing of the LLVM code.
    \item Strict segmentation was enforced between the datasets used for training the embedding models and those used for training and validating neural networks.
  \end{itemize}

\section{Design and Implementation}
\begin{figure}[!htbp]
    \centering
    \includegraphics[width=1\linewidth]{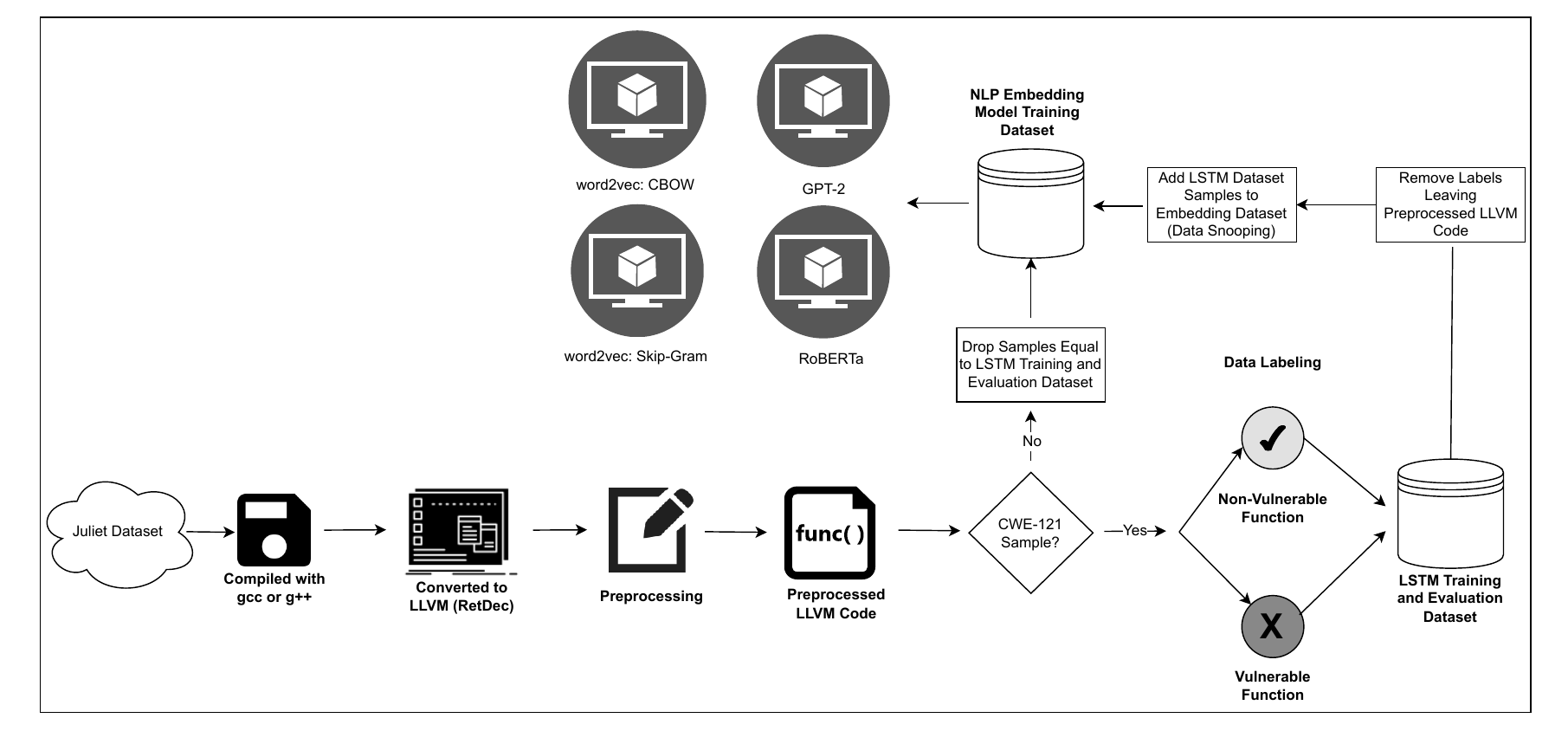}
    \caption{Building Embedding Models with Data Snooping}
    \label{fig:steps1-4}
\end{figure}

\begin{figure}[!htbp]
    \centering
    \includegraphics[width=1\linewidth]{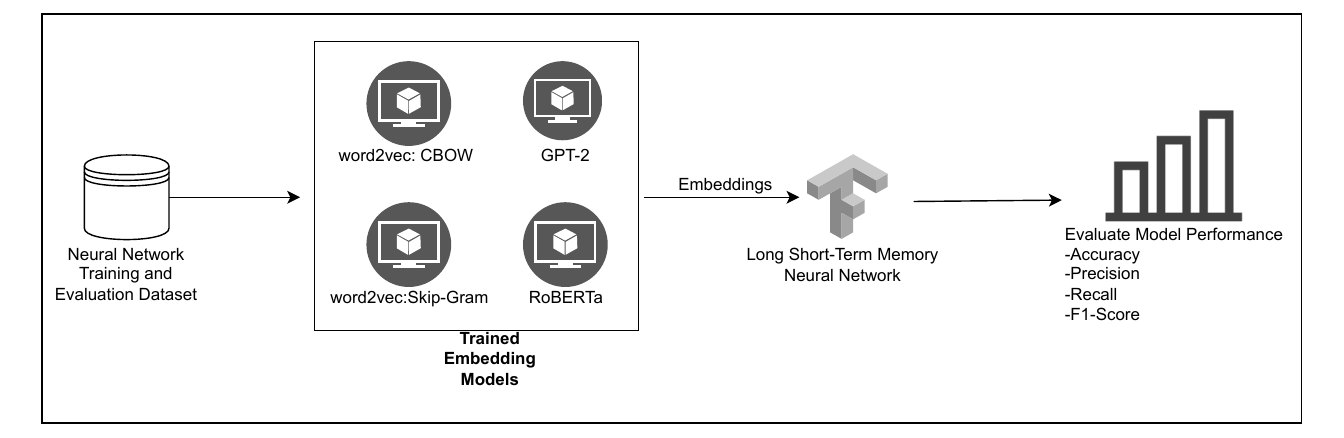}
    \caption{Training Neural Networks to Detect Buffer Overflows}
    \label{fig:steps5-6}
\end{figure}

The current study investigates the effects of test snooping, specifically at the embedding model layer, within the context of the findings in \cite{mccully1} and \cite{mccully2024}. In these previous studies,  LLVM code was provided to multiple embedding models to learn the semantics of the code. After these embedding models were trained, several long-short-term memory (LSTM) neural networks were generated. In this process, LLVM code samples were fed through the embedding model to generate embeddings, which were then provided to LSTM networks to learn to identify vulnerable code. The current study will introduce data snooping into the methodology used by \cite{mccully1} and \cite{mccully2024} by including test and validation samples from neural networks in the dataset used to train embedding models. The results of models generated with and without data snooping will then be compared. This process is as follows and is visually represented in Figs. \ref{fig:steps1-4} and \ref{fig:steps5-6}.
\begin{enumerate}
  \item All SARD Juliet~\cite{RN89} source code dataset samples were compiled into object files. More information regarding the selection of this dataset is in Section \ref{julietdataset}.
  \item These samples were lifted to LLVM intermediate representation (LLVM IR)~\cite{10.1145/2103656.2103709} using a tool named RetDec~\cite{RN22}. This process is covered in Section \ref{liftedsamples}.
  \item Each LLVM IR function was isolated from each lifted object file, and a series of preprocessing steps were applied to this code. More details regarding this process can be found in Section \ref{preprocessing}.
  \item Juliet samples containing stack-based buffer overflow vulnerabilities (CWE-121) were separated from the other LLVM samples. This subset was used to train LSTM neural networks to identify LLVM code with CWE-121 vulnerabilities.
  \item The dataset for training the embedding models was created by randomly dropping LLVM function samples equal to the number of CWE-121 samples. All the CWE-121 samples were added to the dataset used for training the embedding models, intentionally introducing a data snooping condition. This step and the previous step are covered in Section \ref{DatasetCreation}.
  \item Section \ref{EmbeddingModelCreation} covers the creation of the following embedding models trained using the aforementioned dataset:
  \begin{itemize}
    \item GPT-2
    \item RoBERTa
    \item word2vec: CBOW embedding
    \item word2vec: Skip-Gram embedding
  \end{itemize}
  \item The CWE-121 samples were provided to each embedding model, and these embeddings were used to train LSTM neural networks. Section \ref{sec:lstm} contains more information regarding this step in the process.
  \item The performance of these models, which contained the data snooping flaw, was compared with previous results obtained by \cite{mccully1} and \cite{mccully2024}. These results are in Section \ref{sec:results}.
\end{enumerate}

\subsection{Dataset Selection}\label{julietdataset}
The NIST SARD Juliet dataset~\cite{RN89} served as the base dataset for this experiment. This dataset includes C/C++ code samples with various classes of vulnerabilities, such as use-after-free, null pointer dereference, OS command injection, and stack-based buffer overflows. In addition, the data set contains code samples in which these vulnerabilities have been remediated. This dataset is ideal for training machine learning models to distinguish between vulnerable and non-vulnerable code since it comprises both types of samples. 

\subsection{Compiled Code Samples}\label{liftedsamples}
All code samples in the Juliet dataset were compiled into object files on a Kali Linux system using gcc/g++. These object files were lifted to LLVM IR using the RetDec tool~\cite{RN22}. The samples were also compiled on a Windows 10 system, but manual analysis of the lifted LLVM code revealed that RetDec was stripping calls to the standard C library from the lifted code. Since this experiment aimed to train machine learning models to distinguish between vulnerable and non-vulnerable code, the LLVM samples from Windows were excluded.

\subsection{LLVM Preprocessing}\label{preprocessing}
The LLVM functions underwent a series of preprocessing steps in which the code was standardized to eliminate unique characteristics that were not useful for learning code-level semantics. For example, unique function names were renamed to a form that captured functional sequences. Thus, function names such as Unique\_Function\_Name were renamed to func\_x, where x is the sequence in which the function was used within the processed code. Further information regarding the steps used for preprocessing is located in \cite{mccully1} and \cite{mccully2024}. 

\subsection{Embedding and Neural Network Dataset Creation}\label{DatasetCreation}
The preprocessed LLVM samples were divided into two datasets. The first dataset is designed to train a neural network to detect CWE-121 vulnerabilities. Thus, samples containing these vulnerabilities and corresponding remediated versions were isolated from the rest of the LLVM samples. The second dataset contained the rest of the functions preprocessed in step \ref{preprocessing}.

At this stage of the methodology, a data snooping design flaw was introduced into the experiment. To mirror the size of the dataset used in \cite{mccully1} and \cite{mccully2024}, random samples equal in number to those used to train neural networks were removed from the embedding model dataset. Then, the preprocessed samples initially set aside for neural network training were then reintroduced into the embedding model training dataset. This approach ensured that differences in neural network metrics could be more accurately attributed to data snooping rather than changes in the size of the embedding model dataset. The number of samples used in each dataset is in Table \ref{tab:EmbeddingTrainingDataset}.

\begin{table}[!htbp]
\caption{Embedding and LSTM Training Datasets}
\label{tab:EmbeddingTrainingDataset}
\centering
{%
\begin{tabular}{|l|c|c|}
\hline
& \multicolumn{2}{c|}{\textbf{Training (size)}}  \\ \cline{2-3}
\textbf{Dataset Purpose} & LLM & LSTM \\ \hline

Initial Dataset Size	& 48,157 & \makecell{3,802\\ \Xhline{0.5pt} Clean: 2,386\\ Vuln: 1,416} \\ \hline
Post Sample Dropping	& 44,355 & NA \\ \hline
Final Dataset Sizes	& 48,157 & \makecell{3,802\\ \Xhline{0.5pt} Clean: 2,386\\ Vuln: 1,416} \\ \hline
\end{tabular}%
}
\end{table}

\subsection{Embedding Model Creation}\label{EmbeddingModelCreation}
The creation of each embedding model was identical to the processes articulated in \cite{mccully1} and \cite{mccully2024}. Interestingly, the datasets used to train the embedding models resulted in slight gains in the ability of these models to learn the semantics of the LLVM language, as illustrated in Table \ref{tab:Bertrobertaloss}. 
\begin{table}[!htbp]
\caption{Five Last Loss Scores by Validation Step}
\label{tab:Bertrobertaloss}
\resizebox{\columnwidth}{!}{%
\begin{tabular}{|c|cc|cc|cc|cc|}
\hline
\textbf{Step} & \multicolumn{2}{c|}{\makecell{\textbf{RoBERTa}\\\textbf{without Data Snooping}}} & \multicolumn{2}{c|}{\makecell{\textbf{RoBERTa}\\\textbf{with Data Snooping}}}&\multicolumn{2}{c|}{\makecell{\textbf{GPT-2}\\\textbf{without Data Snooping}}} & \multicolumn{2}{c|}{\makecell{\textbf{GPT-2}\\\textbf{with Data Snooping}}} \\
& Training & Validation & Training & Validation& Training & Validation & Training & Validation\\
50,000 & 2.558400 & 2.494601 & 2.551800 & 2.475560 & 0.164600 & 0.132824 & 0.164200 & 0.132083\\
51,000 & 2.557400 & 2.488815 & 2.544400 & 2.474171 & 0.163700 & 0.132133 & 0.164000 & 0.131529\\
52,000 & 2.559900 & 2.485950 & 2.542900 & 2.473276 & 0.162800 & 0.132173 & 0.162300 & 0.131138\\
53,000 & 2.548000 & 2.487699 & \cellcolor{lgreen}2.540800 & \cellcolor{lgreen}2.471620 & 0.162100 & 0.131591 & 0.161700 & 0.130606\\
54,000 & \cellcolor{lgreen}2.557500 & \cellcolor{lgreen}2.483047 & 2.536500 & 2.473666 & \cellcolor{lgreen}0.163200 & \cellcolor{lgreen}0.131189 & \cellcolor{lgreen}0.161800 & \cellcolor{lgreen}0.130531 \\
\hline
\end{tabular}%
}
\end{table}

\subsection{LSTM Neural Network Creation}\label{sec:lstm}
Seven LSTM neural networks were generated using embeddings from each embedding model. Since, without fail, the top-performing neural networks were generated using embedding models with unfrozen layers, all models in the current research were also built using the same. The models were built using the following hyperparameters:
\begin{itemize}
    \item Stochastic Gradient Descent (SGD) optimizer with a learning rate and momentum of 0.01.
    \item SGD optimizer with a learning rate of 0.0001 and momentums of 0.01, 0.001, and 0.0001.
    \item Adam optimizer with initial learning rates of 0.01, 0.001, and 0.0001
\end{itemize}

The embeddings were fed into two hidden LSTM layers, each with 128 neurons and a 20\% dropout rate to help prevent overfitting. The output layer consisted of a single neuron to classify whether each sample contained a CWE-121 vulnerability. Each model was trained over 50 epochs. The dataset was randomly split into training and validation sets, with 3,041 samples (80\%) used for training and 761 samples (20\%) reserved for validation.

\section{Results}\label{sec:results}
\subsection{word2vec: Skip-Gram Metrics}
In \cite{mccully1}, the best-performing model used embeddings from a word2vec Skip-Gram model, achieving a high accuracy of 92.0\% and an F1-score of 89.6\%. However, with data snooping present, this model achieved a top accuracy of 90.5\% and an F1-score of 87.0\%. Overall, the differences in the loss scores, accuracy, precision, recall, and F1 scores were relatively minor. The notable exception to this trend was significant improvements in all metrics when the model was trained using the Adam optimizer with a learning rate of 0.01. The measurable differences that occurred when data snooping was introduced into the experiment are shown in Table \ref{tab:skipPerf}.

\begin{table}[!htbp]
\caption{word2vec: Skip-Gram Metrics with Data Snooping}
\label{tab:skipPerf}
\resizebox{\columnwidth}{!}{%
\begin{tabular}{|l|c|c|c|c|c|c|}
\hline
\makecell[l]{\textbf{Hyper-}\\ \textbf{parameters}} & \textbf{Epoch} & \textbf{Loss} & \textbf{Accuracy} & \textbf{Precision} & \textbf{Recall} &\makecell[l]{\textbf{F1-} \\ \textbf{Score}} \\
\makecell[l]{SGD\\ -LR: 0.01\\ -Mom: 0.01} & 36(-9) & \cellcolor{lgreen}0.3589 (-0.0074) & 80.9\% (±0\%) & \cellcolor{lred}66.7\% (-1.5\%) & \cellcolor{lgreen}96.1\% (+6.1\%) & \cellcolor{lgreen}78.7\% (+1.1\%) \\ \hline

\makecell[l]{Adam\\ -LR: 0.01} & 46(+11) & \cellcolor{lgreen}0.3037 (-0.1090) & \cellcolor{lgreen}89.0\% (+7.9\%) & \cellcolor{lgreen}80.6\% (+11.5\%) & \cellcolor{lgreen}92.1\% (+4.6\%) & \cellcolor{lgreen}86.0\% (+8.8\%) \\ \hline

\makecell[l]{Adam\\ -LR: 0.001} & 48(+5) & \cellcolor{lred}0.2172 (-0.0324) & \cellcolor{lred}90.5\% (-1.5\%) & \cellcolor{lgreen}87.6\% (+2.4\%) & \cellcolor{lred}86.4\% (-8.2\%) & \cellcolor{lred}87.0\% (-2.6\%) \\ \hline

\makecell[l]{Adam (DS)\\ -LR: 0.0001} & 41(-7) & \cellcolor{lred}0.3326 (+0.0044) & \cellcolor{lred}83.6\% (-0.4\%) & \cellcolor{lred}71.9\% (-0.2\%) & \cellcolor{lred}90.7\% (-1.1\%) & \cellcolor{lred}80.2\% (-0.6\%) \\ \hline
\end{tabular}%
}
\end{table}

\subsection{word2vec: CBOW Metrics}
In \cite{mccully1}, the best performing neural network trained with a word2vec: CBOW embedding model achieved a high accuracy of 87.5\% and an F1-score of 84.3\%. Interestingly, when data snooping was present, the top-performing neural network achieved a similarly high accuracy of 87.9\% and an F1-score of 84.2\%. These almost identical scores indicate that data snooping did not play a significant factor in the neural network's ability to generalize LLVM samples containing CWE-121 vulnerabilities. Table \ref{tab:cbowPerf} shows the overall differences between neural networks trained using embeddings with and without data snooping present.  

\begin{table}[!htbp]
\caption{word2vec: CBOW Metrics with Data Snooping}
\label{tab:cbowPerf}
\resizebox{\columnwidth}{!}{%
\begin{tabular}{|l|c|c|c|c|c|c|}
\hline
\makecell[l]{\textbf{Hyper-}\\ \textbf{parameters}} & \textbf{Epoch} & \textbf{Loss} & \textbf{Accuracy} & \textbf{Precision} & \textbf{Recall} &\makecell[l]{\textbf{F1-} \\ \textbf{Score}} \\
\makecell[l]{SGD\\ -LR: 0.01\\ -Mom: 0.01} & 44(+1) & \cellcolor{lred}0.3423 (+0.0006) & \cellcolor{lgreen}83.6\% (+0.6\%) & \cellcolor{lgreen}72.0\% (+0.2\%) & \cellcolor{lgreen}90.3\% (+1.8\%) & \cellcolor{lgreen}80.1\% (+0.8\%) \\ \hline

\makecell[l]{SGD\\ -LR: 0.0001\\ -Mom: 0.01} & 50(±0) &  \cellcolor{lred}0.6121 (+0.0201) & \cellcolor{lred}67.8\% (-2.0\%) & \cellcolor{lgreen}72.4\% (+4.5\%) & \cellcolor{lred}19.7\% (-13.6\%) & \cellcolor{lred}31.0\% (-13.7\%) \\ \hline

\makecell[l]{SGD\\ -LR: 0.0001\\ -Mom: 0.001} & 37(-5) & \cellcolor{lgreen}0.6163 (-0.0083) & \cellcolor{lgreen}69.3\% (+0.3\%) & \cellcolor{lgreen}74.7\% (+10.5\%) & \cellcolor{lred}24.4\% (-10.4\%) & \cellcolor{lred}36.8 (-8.3\%)\% \\ \hline

\makecell[l]{SGD\\ -LR: 0.0001\\ -Mom: 0.0001} & 48(+11) & 0.\cellcolor{lgreen}6114(-0.0045) & \cellcolor{lgreen}69.8\% (+1.1\%) & \cellcolor{lgreen}67.4\% (+4.0\%) & \cellcolor{lred}34.1\% (-0.7\%) & \cellcolor{lgreen}45.2\% (+0.3\%) \\ \hline

\makecell[l]{Adam\\ -LR: 0.01} & 31(-15) & \cellcolor{lgreen}0.3285 (-0.0246) & \cellcolor{lgreen}84.2\% (+0.8\%) & \cellcolor{lgreen}71.1\% (+0.4\%) & \cellcolor{lgreen}96.1\% (+2.6\%) & \cellcolor{lgreen}81.7\% (+1.1\%) \\ \hline

\makecell[l]{Adam\\ -LR: 0.001} & 46(+12) & \cellcolor{lred}0.2920 (+0.0392) & \cellcolor{lgreen}87.9\% (+0.4\%) & \cellcolor{lgreen}80.7\% (+2.7\%) & \cellcolor{lred}88.2\% (-3.6\%) & \cellcolor{lred}84.2\% (-0.1) \\ \hline

\makecell[l]{Adam\\ -LR: 0.0001} & 46(+1) & \cellcolor{lgreen}0.2987 (-0.0003) & \cellcolor{lred}85.8\% (-0.7\%) & \cellcolor{lred}75.7\% (-1.3\%) & \cellcolor{lred}90.3\% (0.3\%) & \cellcolor{lred}82.4\% (-0.6) \\ \hline
\end{tabular}%
}
\end{table}

\subsection{RoBERTa Metrics}
When data snooping was present, the neural network that achieved the best performance using RoBERTa embeddings achieved a high accuracy of 89.1\% with an F1-Score of 85.4\%. This is slightly better than the model that achieved an accuracy of 88.8\% and an F1-Score of 84.2\%, achieved in \cite{mccully1}. Table \ref{tab:RoBERTaPerf} shows the metrics differences observed in neural networks when data snooping was introduced in the RoBERTa model used to generate embeddings. 

\begin{table}[!htbp]
\caption{RoBERTa Metrics with Data Snooping}
\label{tab:RoBERTaPerf}
\resizebox{\columnwidth}{!}{%
\begin{tabular}{|l|c|c|c|c|c|c|}
\hline
\makecell[l]{\textbf{Hyper-}\\ \textbf{parameters}} & \textbf{Epoch} & \textbf{Loss} & \textbf{Accuracy} & \textbf{Precision} & \textbf{Recall} &\makecell[l]{\textbf{F1-} \\ \textbf{Score}} \\
\makecell[l]{SGD\\ -LR: 0.01\\ -Mom: 0.01} & 34(+3) & \cellcolor{lred}0.3839 (+0.1214) & \cellcolor{lred}85.2\% (-1.4\%) & \cellcolor{lgreen}87.4\% (+5.2\%) & \cellcolor{lred}69.5\% (-11.5\%) & \cellcolor{lred}77.4\% (-4.2\%) \\ \hline

\makecell[l]{SGD\\ -LR: 0.0001\\ -Mom: 0.01} & 32(+15) & \cellcolor{lgreen}0.2447 (-0.0007) & \cellcolor{lgreen}89.0\% (+0.3\%) & \cellcolor{lgreen}81.2\% (+2.2\%) & \cellcolor{lgreen}91.0\% (+3.3\%) & \cellcolor{lred}85.8\% (-0.1\%) \\ \hline

\makecell[l]{SGD\\ -LR: 0.0001\\ -Mom: 0.001} & 45(±0) & \cellcolor{lgreen}0.2578 (-0.0130) & \cellcolor{lred}88.6\% (-0.2\%) & \cellcolor{lred}81.2\% (-6.4\%) & \cellcolor{lgreen}89.6\% (+8.6\%) & \cellcolor{lgreen}85.2\% (+1.0\%) \\ \hline

\makecell[l]{SGD (DS)\\ -LR: 0.0001\\ -Mom: 0.0001} & 48(+1) & \cellcolor{lgreen}0.2377 (-0.0491) & \cellcolor{lgreen}89.1\% (+0.5\%) & \cellcolor{lred}83.8\% (-0.2\%) & \cellcolor{lgreen}87.1\% (+2.2\%) & \cellcolor{lgreen}85.4\% (+0.9\%) \\ \hline
\end{tabular}%
}
\end{table}

\subsection{GPT-2 Metrics}
The best-performing neural networks created using GPT-2 embeddings without data snooping outperformed those with data snooping. Table \ref{tab:GPTPerfUnfrozen} shows that the best-performing model when data snooping was present achieved an accuracy of 92.1\% and an F1-Score of 88.6\%. However, in \cite{mccully2024}, the best performing neural network created without data snooping present achieved an accuracy of 92.5\% and an F1-Score of 89.7\%. 

\begin{table}[!htbp]
\caption{GPT-2 Metrics with Data Snooping}
\label{tab:GPTPerfUnfrozen}
\resizebox{\columnwidth}{!}{%
\begin{tabular}{|l|c|c|c|c|c|c|}
\hline
\makecell[l]{\textbf{Hyper-}\\ \textbf{parameters}} & \textbf{Epoch} & \textbf{Loss} & \textbf{Accuracy} & \textbf{Precision} & \textbf{Recall} &\makecell[l]{\textbf{F1-} \\ \textbf{Score}} \\
\makecell[l]{SGD\\ -LR: 0.01\\ -Mom: 0.01} & 44(-6) & \cellcolor{lred}0.3339 (+0.0290) & \cellcolor{lgreen}89.0\% (+0.4\%) & \cellcolor{lgreen}91.8\% (+7.3\%) & \cellcolor{lred}76.7\% (-7.5\%) & \cellcolor{lred}83.6\% (-0.8\%) \\ \hline

\makecell[l]{SGD\\ -LR: 0.0001\\ -Mom: 0.01} & 46(+2) & \cellcolor{lred}0.1691 (+0.0103) & \cellcolor{lred}91.2\% (-1.3\%) & \cellcolor{lred}89.3\% (-0.9\%) & \cellcolor{lred}86.4\% (-2.8\%) & \cellcolor{lred}87.8\% (-1.9\%) \\ \hline

\makecell[l]{SGD\\ -LR: 0.0001\\ -Mom: 0.001} & 43(+1) & \cellcolor{lgreen}0.1517 (-0.0195) & 92.1\% (±0\%) & \cellcolor{lgreen}94.0\% (+2.4\%) & \cellcolor{lred}83.9\% (-2.5\%) & \cellcolor{lred}88.6\% (-0.3\%) \\ \hline

\makecell[l]{SGD\\ -LR: 0.0001\\ -Mom: 0.0001} & 44 (-2) & \cellcolor{lred}0.2007 (+0.0485) & \cellcolor{lred}91.6\% (-0.5\%) & \cellcolor{lred}87.2\% (-4.1\%) & \cellcolor{lgreen}90.3\% (+3.6\%) & \cellcolor{lred}88.7\% (-0.3\%) \\ \hline

\makecell[l]{Adam\\ -LR: 0.0001} & 10(-7) & \cellcolor{lgreen}0.3033 (-0.0963) & \cellcolor{lgreen}89.6\% (+2.5\%) & \cellcolor{lgreen}81.6\% (+2.5\%) & \cellcolor{lgreen}92.5\% (+4.3\%) & \cellcolor{lgreen}86.7\% (+3.3\%) \\ \hline
\end{tabular}%
}
\end{table}

\subsection{Summary of Results}
Table \ref{tab:BestModels} provides a comparison of the best-performing model in \cite{mccully1} and \cite{mccully2024} with those generated during the current research. As this table shows, the effects of data snooping on the accuracy and F1-score of the best-performing models are quite modest. In the current research project context, data snooping did not directly correlate to a higher ability by the neural networks to classify vulnerable and non-vulnerable code correctly.

\begin{table}[!htbp]
\caption{Best Performing Models}
\label{tab:BestModels}
\centering
\begin{tabular}{|c|cc|cc|}
\hline
\textbf{Model} & \multicolumn{2}{c|}{\textbf{without Data Snooping}} & \multicolumn{2}{c|}{\textbf{with Data Snooping}} \\
& \textbf{Accuracy} & \textbf{F1-Score} & \textbf{Accuracy} & \textbf{F1-Score} \\
\hline
GPT-2 & \cellcolor{lgreen}92.5\% & \cellcolor{lgreen}89.7\% & 92.1\% & 88.6\% \\\hline
Skip-Gram & \cellcolor{lgreen}92.0\% & \cellcolor{lgreen}89.6\% & 90.5\% & 87.0\% \\\hline
RoBERTa & 88.8\% & 84.2\% & \cellcolor{lgreen}89.1\% & \cellcolor{lgreen}85.4\% \\\hline
CBOW & 87.5\% & 84.3\% & \cellcolor{lgreen}87.9\% & \cellcolor{lgreen}84.2\% \\
\hline
\end{tabular}%
\end{table}

\section{Other Data Snooping Categories}\label{sec:tempselsnooping}
\subsection{Temporal and Selective Snooping}
Although this research focuses on the effects of test snooping introduced in embedding models, two additional categories of data snooping are worth considering. \textit{Temporal snooping} is a second form of data snooping, which occurs when the researchers ignore time dependencies within the datasets used in the research~\cite{arp2022and}. The subcategories of this form of data snooping are defined in Table \ref{tab:temporal_snooping}.
\begin{table}[h!]
\centering
\caption{Temporal Snooping Subcategories (\textit{Source:} \cite{arp2022and})}
\begin{tabular}{|p{0.2\columnwidth}|p{0.70\columnwidth}|}
\hline
\textbf{Category} & \textbf{Description} \\ \hline
Time Dependency & The samples used to train and validate the model are time-dependent, meaning that features included may not be available or relevant in future scenarios, potentially limiting the model's ability to detect similar issues as data evolves. \\ \hline
Aging Datasets & The use of publicly available datasets widely used in prior research can introduce bias, as researchers may inadvertently incorporate prior knowledge from previous analyses. \\ \hline
\end{tabular}
\label{tab:temporal_snooping}
\end{table}

The last category of data snooping is \textit{selective snooping}~\cite{arp2022and}, which can be introduced during dataset cleaning~\cite{arp2022and}. The subcategories for this type of snooping are in Table \ref{tab:selective_snooping}.

\begin{table}[h!]
\centering
\caption{Selective Snooping Subcategories (\textit{Source:} \cite{arp2022and})}
\begin{tabular}{|p{0.2\columnwidth}|p{0.70\columnwidth}|}
\hline
\textbf{Category} & \textbf{Description} \\ \hline
Cherry-picking & Samples are filtered or cleaned based on information not realistically available during actual use. \\ \hline
Survivorship Bias & Samples are filtered based on conditions that exclude certain data, which can introduce survivorship bias by only including data that meets specific criteria. \\ \hline
\end{tabular}
\label{tab:selective_snooping}
\end{table}

\subsection{Temporal and Selective Snooping in Context of Previous Research}
\subsubsection{Temporal Snooping}
\cite{mccully1} and \cite{mccully2024} used the Juliet 1.3 dataset~\cite{RN11}, and the neural networks in the research are designed to detect stack-based buffer overflows (CWE-121) in lifted code. Determining if these code samples have a time dependency is challenging. Buffer overflows have been successfully exploited for decades, with instances dating back to worms as early as 1988~\cite{4017740}. The concept behind exploiting CWE-121 vulnerabilities is well-established and is unlikely to change over time. However, new patterns that lead to buffer overflow vulnerabilities may emerge. For example, one common attack pattern leading to CWE-121 vulnerabilities involves insecure functions in the C/C++ programming languages, such as \texttt{strcpy} and \texttt{fgets}~\cite{larochelle2001statically}. 

Other programming languages may also introduce CWE-121 vulnerabilities in the future. However, unforeseen future conditions that may result in new vulnerability classifications should not be considered a design flaw. Classification models will always need 
updates to accommodate new data variants.

\cite{mccully1} and \cite{mccully2024} used the Juliet 1.3 C/C++ source code dataset for their research. Although some researchers~\cite{white2000reality,arp2022and} may argue this could introduce data snooping because other researchers have used the dataset for similar purposes, the use of standard datasets for model evaluation is a standard practice among researchers for classification tasks. For example, comparing CNN capabilities using the same dataset is a standard process~\cite{russakovsky2015imagenet}. Although it is critical that baselines properly represent the conditions it proposes to represent~\cite{8586840}, a lack of baselines makes meaningful comparison almost impossible. The Juliet dataset that \cite{mccully1} and \cite{mccully2024} used is maintained through the NIST SARD project~\cite{RN11} and is widely used for the task of research involving the identification of vulnerabilities in compiled code. Some of the previous studies that have used this dataset are \cite{RN50,RN43,RN67,RN56,RN47,RN46,RN55,schaad2022deeplearningbased}. Thus, rather than being a design flaw, using this dataset only strengthens the research by providing a baseline already used by other researchers.

\subsubsection{Selective Snooping}
When building the dataset used for training the neural networks in \cite{mccully1} and \cite{mccully2024}, samples with tokens more than 2048 were removed from the dataset. This choice was made to avoid resource issues when training each neural network. This choice resulted in the loss of 99 of the 3,901 samples in this dataset (approx. 2.5\% loss). The researchers deemed this an acceptable loss as the low number of excluded samples did not introduce survivorship bias significant enough to justify the added complexity required to retain them.

Another example of survivorship bias in \cite{mccully1} and \cite{mccully2024} could have arisen by removing samples where calls to the standard C library were stripped when lifted to LLVM. However, including these samples would have been detrimental to the experiment. Training a model with samples that lack vulnerabilities but instructing the model to classify them as vulnerable would compromise the experiment's objectives. Therefore, removing these samples was a justified and necessary decision.

The final example of survivorship bias in \cite{mccully1} and \cite{mccully2024} would be that only source-code samples that could be successfully compiled were lifted to LLVM for processing. However, this limitation was a necessary aspect of the research methodology. It would not have been possible to train a model on lifted code, if the source code could not be compiled first.

\section{Discussion}
\subsection{Embedding Model Consistency}
Notably, the ranking of the best-performing models based on specific embedding methods remained consistent, regardless of the presence of data snooping. In \cite{mccully2024}, the neural networks trained with embeddings were ranked as follows: GPT-2, Skip-Gram, RoBERTa, and CBOW. Remarkably, this same hierarchy persisted even with data snooping, reinforcing the robustness of GPT-2 embeddings as the most effective for detecting vulnerable LLVM code. This consistency among the experimental conditions strengthens the conclusion that the GPT-2 model produces superior embeddings for training neural networks, as demonstrated in \cite{mccully2024}.

\subsection{Embedding Model Resiliency}
The similar model metrics observed with and without data snooping suggest a notable resilience in the models developed in \cite{mccully1} and \cite{mccully2024}. Data snooping is generally expected to exaggerate neural network performance. However, the absence of this boost may indicate robustness to data contamination and noise, which underscores the reliability of the models even under less-than-ideal conditions.

\subsection{Low Risk of Model Overfitting}
The observed constancy between the models generated using embeddings with and without data snooping may indicate that the models built in \cite{mccully1} and \cite{mccully2024} are not prone to overfitting. This means that instead of fitting the models to specific instances of CWE-121 vulnerabilities, it can generalize the conditions that result in the vulnerability. This is an advantageous characteristic of these models because it enables the model to identify code sequences that may lead to an instance of a CWE-121 vulnerability even if it has not encountered the exact code sequence that resulted in that condition.  

\section{Related Work}\label{relatedworks}
\cite{white2000reality} focused their research on the challenges of temporal snooping, specifically the repeated use of datasets. In the context of time-series data for financial forecasting, they highlight several studies addressing this issue~\cite{white2000reality}. The authors introduce methods to determine whether the observed differences between models result from the reuse of datasets or reflect genuine differences within the underlying models, allowing researchers to interpret their findings confidently~\cite{white2000reality}. \cite{romano2005stepwise} also addresses the challenges of temporal snooping in the context of time series data used for financial analysis. However, this paper focuses on techniques designed to assess whether observed differences in model performance can be attributed to chance, particularly in scenarios involving multiple simultaneous tests~\cite{romano2005stepwise}. \cite{neuhierl2011data} applied the techniques proposed by \cite{white2000reality,romano2005stepwise,hsu2010testing,hansen2005test} to evaluate a series of market-timing rules. These tests aimed to determine whether the results were statistically significant after adjustment for temporal snooping~\cite{neuhierl2011data}.

Building on the work of \cite{amiri2013data}, \cite{10379028} introduces a selective data snooping technique integrated with a generalized error-in-variables model to improve pose estimation accuracy. The approach achieves this goal by identifying and removing data points contributing to gross errors~\cite{10379028}.

\section{Future Research}
The dataset used to train the embedding models was limited in size, containing approximately 48,000 samples. Similarly, the datasets for training and evaluating the neural networks were also limited in size. Models trained with larger or smaller datasets may exhibit different degrees of susceptibility to performance variations. Future research will explore the effects of the size of the dataset and the impact of data snooping on models trained with varying sample sizes.

It was assumed that including the neural network training and validation sets in the dataset used to train the embedding model would result in a more significant increase in model performance. However, this was not observed in this experiment. Thus, additional research would need to be completed to further investigate why no performance increase was observed. As part of the methodology, a random selection of samples in the embedding model training dataset was replaced with those in the neural network training dataset. Some randomly dropped samples may have contained hidden attributes that helped the embedding models better represent conditions in the LLVM code that resulted in a vulnerable condition (e.g., token sequence sizes, LLVM keywords, specific token sequences, etc.). Future research should investigate potential hidden features within the code and the impact these features may have on generated embeddings.

\section{Conclusion}
The results of this experiment were surprising. The researchers initially assumed that combining the dataset for training neural networks with the dataset for training embedding models would consistently enhance neural network performance. However, the absence of such improvements suggests that either mixing the datasets had minimal impact on the base models or that hidden features in some of the randomly dropped samples used in the original models prevented the new models from reaching their full potential.

This study reinforces that neural networks using GPT-2 embeddings consistently outperform those using other embedding models, validating previous conclusions from \cite{mccully2024} under both data and non-data snooping conditions. These findings suggest a notable robustness in GPT-2 embeddings, highlighting their potential for practical applications in vulnerability detection, even under less-than-ideal conditions.

Future research should investigate the hidden underlying features and explore more extensive or more varied datasets to better understand the limitations and strengths of embedding models in vulnerability detection. Additionally, examining the effects of data snooping in other cybersecurity-related machine learning tasks may provide further insights into its impact across various cybersecurity applications.
%
%

\printbibliography

\end{document}